\documentclass[prb,reprint,floatfix,nobalancelastpage]{revtex4-2} 

\usepackage{amsmath}
\usepackage{amsfonts}
\usepackage{graphicx}

\usepackage[english]{babel}
\usepackage[utf8]{inputenc}
\usepackage[T1]{fontenc}
\usepackage{longtable}
\usepackage[bookmarks,colorlinks=false]{hyperref}

\begin{document}

\title{Black Hole Flyby}

\author{Sebastian J. Szybka}
\affiliation{Jagiellonian University and Copernicus Center for Interdisciplinary Studies}


\begin{abstract}
We calculate the minimum distance at which one may approach a black hole in a free flyby. It corresponds to $r=4m$ for the Schwarzschild black hole for a probe that was non-relativistic at infinity. The problem is formulated in a way that is useful for teaching introductory general relativity.
\end{abstract}

\maketitle

\section{Introduction}

Recently, the first-ever direct image of a black hole has been presented by the Event Horizon Telescope collaboration \cite{eht1,eht2}. The image shows the supermassive black hole at the centre of the galaxy M87. A central dark region---the black hole shadow---is surrounded by the bright ring. The existence of the shadow was highly anticipated. The size of this region for a static black hole was predicted by Synge \cite{synge}. The predominantly straightforward calculations involve photons' trajectories and can be conducted at the undergraduate level as has been presented in the recent textbook by Chruściel \cite{chrusciel}. Surprisingly, the importance of the black hole shadow has been largely overlooked in the textbooks published before the Event Horizon Telescope observations with notable exceptions of the books by Frolov and Zelnikov \cite{frolov} and Misner, Thorne, and Wheeler \cite{gravitation}. In this article, we point out that for massive particles there exists an analogous basic problem that has not been brought to full light in books on general relativity.

Imagine that you want to send a probe towards a black hole and intercept it after a flyby. How close to the black hole can the probe go without using propulsion? How large will the time dilation effect be relative to stationary observers far away from the black hole? Answers to these questions follow straightforwardly from formulas presented in textbooks on general relativity, but, to our knowledge, the problem has not been directly discussed in a simple manner before. The solution is elegant and provides an instructive exercise for general relativity students. 

The Schwarzschild spacetime was discovered more than one hundred years ago \cite{gron}. Over these years, a vast body of literature on the Schwarzschild geodesics has emerged. The early efforts \cite{droste,hagihara,darwin,mielnik} culminated in the Chandrasekhar book \cite{chandra}. A thorough analysis of geodesics in the Schwarzschild spacetime may be found also in many other sources, e.g.,\ Ref.\ \cite{hioe} (null and timelike geodesics) and Ref.\ \cite{munoz} (null geodesics). A particular type of orbits we are interested in was studied in Ref.\ \cite{emanuele} (for some related exercises see Sec.\ $10$ in Ref.\ \cite{moore}). Our work does not extend this knowledge with new mathematical results, but makes what is known more accessible at the introductory level. Our narrative should build readers' intuition about strongly relativistic systems that these days are more relevant than ever.

\section{Black hole shadow}

Before we formulate our problem in detail, an explanatory note on a black hole shadow is needed.

A black hole is a region of space-time of no escape. Whatever falls in cannot return. The boundary of this region is called an event horizon. 

In this work, we consider a static black hole which is described by the Schwarzschild solution. The line element is 
\begin{equation}\label{metric}
ds^2=-\left(1-\frac{2m}{r}\right)dt^2+\left(1-\frac{2m}{r}\right)^{-1}dr^2+r^2d\Omega^2\;,
\end{equation}
where $m>0$ is the mass of the black hole, $d\Omega^2=d\theta^2+\sin^2\theta d\varphi^2$ is the metric on the unit two-dimensional sphere, and $-\infty<t<\infty$, $r>2m$, $0<\theta<\pi$, $0<\varphi<2\pi$. 

We use geometric units $G=c=1$. Usage of this system of units simplifies formulas, but it might be initially confusing, e.g.,\ in geometric units apparently different physical quantities like time, space, and mass are measured in meters. A conversion factor from/to geometric units can be found easily with the help of dimensional analysis. We show how to do that at the beginning of Sec.~\ref{examples}.

A classical black hole does not radiate. Since it captures light, it obstructs the view of its accretion disk and background stars. A black hole is visible via its shadow. Naively, one may think that the size of the region hidden from view is related to the size of the event horizon $r=2m$. A more educated guess would relate the size of the central dark region to the inner edge of the accretion disk which illuminates the black hole. This inner edge corresponds roughly to the innermost stable circular orbit $r=6m$. Neither guess is correct. In order to determine the size of the black hole shadow, it is convenient to formulate the problem as follows.

\begin{figure}[t!]
\centering
\includegraphics[scale=0.65]{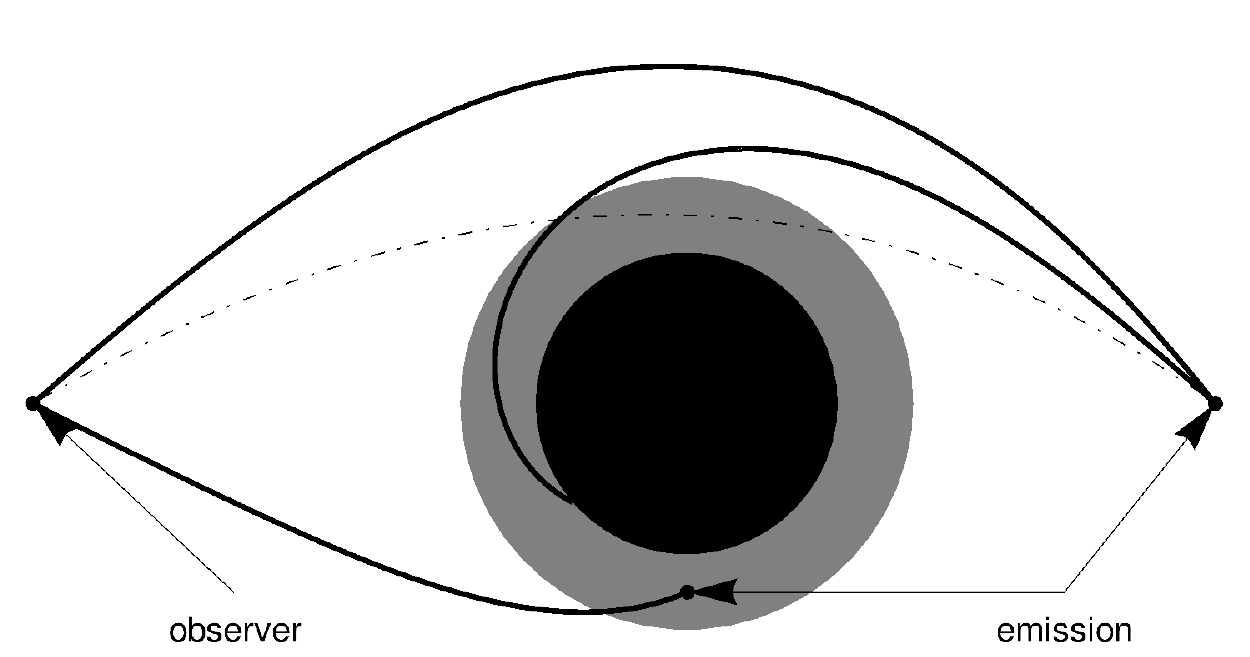}
\caption{Examples of photons' trajectories in the Schwarzschild spacetime (solid). The dotted-dashed trajectory is inconsistent with the geodesic equation. The black hole corresponds to a black disk with a radius $r=2m$. The larger gray disk has a radius $r=3m$.}
\label{figI}
\end{figure}

We consider a photon emitted from a large distance towards a black hole (the top trajectory in Fig.~\ref{figI}). The photon passes near the black hole and flies away from it approaching a distant observer. One can easily calculate the minimum distance at which such a photon passed the black hole. It is larger than $r=3m$. The coordinate radius $r=3m$ corresponds to the unstable circular photon orbit encircling the black hole. All photons that were radiated outside of the sphere $r=3m$ and crossed $r=3m$ are swallowed by the black hole. Of course, the distant observer can still see photons that were emitted above the event horizon in the region $2m<r\leq 3m$. However, since the accretion disk---the dominant source of radiation---lies outside of the sphere $r=3m$, then the size of the black hole shadow will correspond to $r=3m$. The observational importance of the circular photon orbits $r=3m$ might be surprising, because these orbits are unstable.

The angular size of the black hole shadow for a distant observer is approximately equal to $6\sqrt{3}\,m/r$, where $r\gg 2m$ is the distance to the black hole \cite{synge,chrusciel}. We point out that due to relativistic effects it does not correspond to $2\cdot3m/r=6m/r$ as might be naively expected.

\section{Close encounter with black hole}

For massive test particles, one may study an analogous problem. We consider a probe which is freely falling towards a black hole. The probe collects data at the closest approach and stays on an asymptotically circular orbit (endlessly circling the black hole) or flies away to be intercepted by a distant stationary observer. How close could a black hole be approached in such a flyby without using propulsion?

The mass of the probe is many orders of magnitude smaller than that of the black hole. Therefore, the probe may be treated as a massive test particle with the four-velocity $u^\alpha$. The probe is freely falling, so $u^\alpha$ is tangent to a timelike geodesic. For the sake of clarity, we rederive below the equations that govern timelike geodesics in the Schwarzschild spacetime.

The components of the Schwarzchild metric (\ref{metric}) do not depend on the coordinates $t$ and $\varphi$, thus the basic considerations imply that the covariant components of the four velocity $u_t$, $u_\varphi$ are conserved along geodesics. Hence, $u_t=-e$, $u_\varphi=l$, where $e$, $l$ are constants \cite{footnote1}. The symmetry of the problem implies that, without loss of generality, we may assume $\theta=\pi/2$ for the entire trajectory, so $u^\theta=0$. Using $u_\alpha=g_{\alpha\beta}u^\beta$ the covariant components of the four-velocity along the geodesic can therefore be written as
\begin{equation}\label{ucc}
u_\alpha=(-e,g_{rr}u^r,0,l)\;.
\end{equation}
Since $u^\alpha$ is a future-oriented timelike vector, the constant $e$ must satisfy $e>0$.
An asymptotic flatness of the Schwarzschild spacetime may be used to find a physical interpretation of $e$ as follows. We consider a probe that was initially far away from the black hole moving with velocity $v_\infty$ relative to distant stationary observers. There we have $u^t=\gamma_\infty=-u_t=e$, where $\gamma_\infty=1/\sqrt{1-(v_\infty/c)^2}$ is the Lorentz factor of the probe. Therefore, the constant $e$ corresponds to ``energy per mass.'' (For particles that are not gravitationally bound, $e\geq 1$.) Similarly, one may show that $l$ is ``an angular momentum per mass.'' The situation we are interested in excludes radial orbits, so $l\neq 0$ and without loss of generality we assume $l>0$. 

The components $u_t=-e$, $u_\varphi=l$, $u^\theta=0$ are first integrals of the geodesic equation. The fourth first integral is given by the normalization condition of the four-velocity $u^\alpha$,
\begin{equation}\label{nc}
u_\alpha u^\alpha=-1\;.
\end{equation}
Using these four constants of motion, we reduced the geodesic equation to the first-order system of ordinary differential equations. Since the components of the metric (\ref{metric}) depend on $r$ only, it is convenient to introduce a pseudopotential and apply techniques known from classical mechanics. We substitute Eq.\ (\ref{ucc}) into Eq.\ (\ref{nc}), where $u^\alpha$ is calculated as $u^\alpha=g^{\alpha\beta}u_\beta$. The metric (\ref{metric}) is diagonal, thus its contravariant components $g^{\alpha\beta}$ can be found easily. After some algebraic manipulations we obtain
\begin{equation}\label{maineq}
E=\frac{e^2-1}{2}=\frac{1}{2}{(u^r)}^2+V_l(r)\;,
\end{equation}
where $E>-1/2$ is a new auxiliary constant, and
\begin{equation}\label{pot}
V_l(r)=-\frac{m}{r}+\frac{l^2}{2r^2}-\frac{m l^2}{r^3}\;.
\end{equation}
To sum up, we have reduced the geodesic equation to Eq.\ (\ref{maineq}) for $u^r=dr/d\tau$ and the equations $u^t=dt/d\tau=u_t g^{tt}=e/(1-2m/r)$, $u^\theta=d\theta/d\tau=0$, and $u^\varphi=d\varphi/d\tau=u_\varphi g^{\varphi\varphi}=l/r^2$, where $\tau$ is a proper time of the probe along the geodesic. 

There is a formal correspondence between Eq.\ (\ref{maineq}) and the Newtonian problem of a body moving in a potential field. In this analogy, the constant $E$ plays a role of total energy per mass of the body. On the right-hand-side of Eq.\ (\ref{maineq}), one recognizes an analog of the kinetic term and the potential. This motivates us to refer to $V_l(r)$ as pseudopotential. It is instructive to note that the first two terms in Eq.\ (\ref{pot}) are the same as in the Newtonian version of our problem (gravitational attraction and the angular momentum barrier). The last term $\sim 1/r^3$ is purely relativistic. It is negative for $l\neq 0$ and dominates the pseudopotential for small $r$: A black hole, in contrast to a Newtonian point mass, can capture test particles even if they have non-zero angular momentum.

\begin{figure}[t!]
\centering
\includegraphics[scale=0.65]{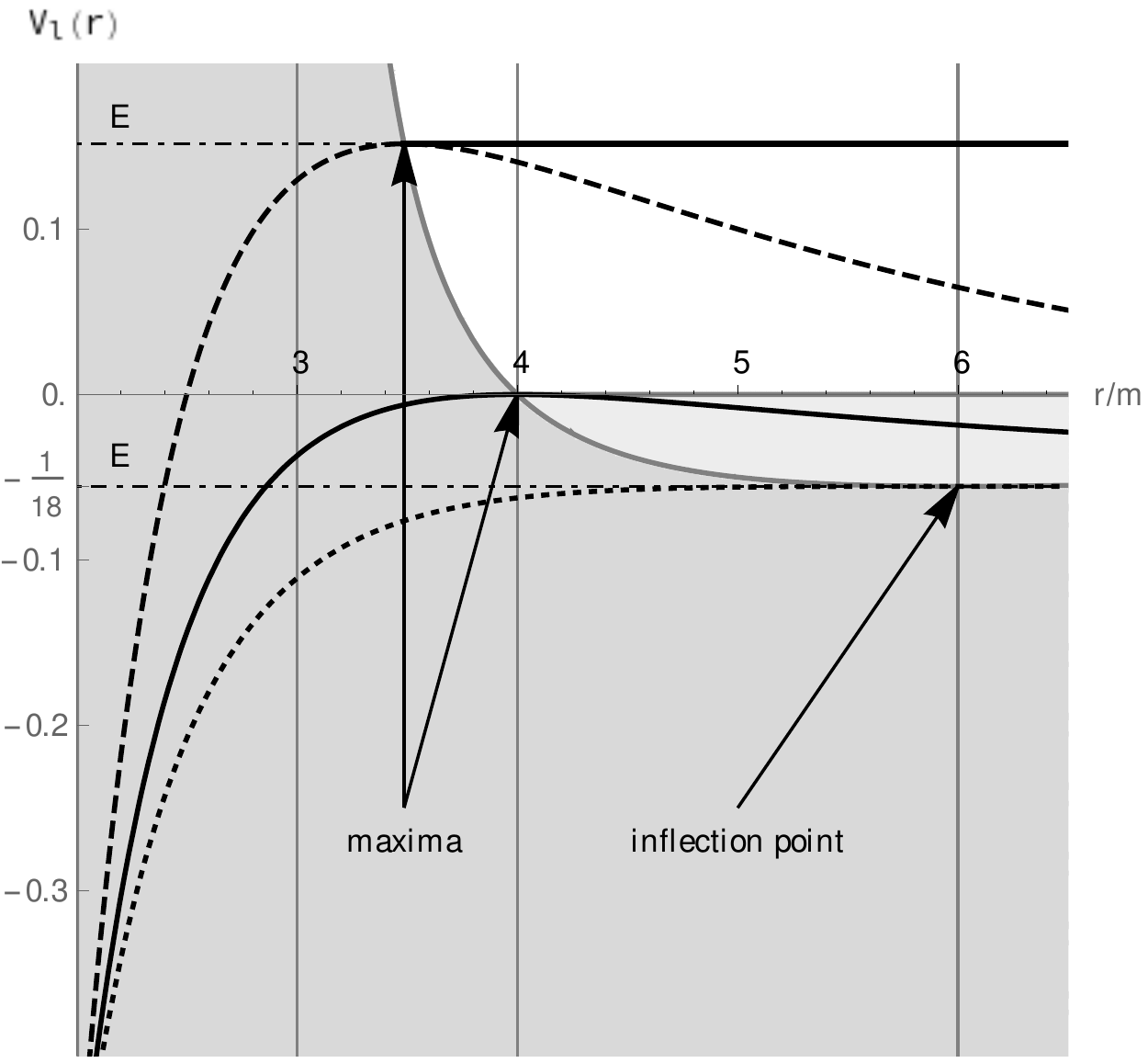}
\caption{The pseudopotential for the angular momentum per mass equal to $l=\sqrt{12}m$ (dotted), $l=4m$ (solid), $l=5m$ (dashed) (massive particles). For a fixed value of $E$ the probe cannot enter the ``forbidden region'' (the gray color). The light gray (above the forbidden gray region and below the line $V_l=0$) indicates the zone which may be penetrated by gravitationally bound particles that do not fall into the black hole. The boundary of the forbidden region asymptotes to $r=3m$ as $E\rightarrow +\infty$, where $r=3m$ coincides with unstable circular photon orbit related to the boundary of the black hole shadow. $E=-1/18$ with $l=l_{crit}(E)=\sqrt{12}m$ corresponds to the innermost stable circular orbit and the inflection point $r=6m$. The probe that was non-relativistic at infinity ($E=0$) and which does not fall into the black hole cannot cross the sphere \mbox{$r=4m$}.}
\label{fig0}
\end{figure}

\subsection{A minimum distance to a black hole}\label{sec:min}

The radial coordinate $r$ of the probe decreases initially, so $u^r<0$. At the minimum distance to a black hole, the component $u^r$ changes sign going smoothly through zero or approaches zero asymptotically. It is clear from the shape of the pseudopotential (Fig.~\ref{fig0}) that for a fixed $l$ the minimum distance coincides with the innermost local maximum of the pseudopotential. The condition for the extrema of $V_l'(r_\pm)=0$ leads to the quadratic equation
\begin{equation}\label{Vp}
mr_\pm^2-l^2r_\pm+3ml^2=0\;,
\end{equation}
with the solutions (this explicit form of the solutions is not necessary to solve the main problem)
\begin{equation}
r_\pm=\frac{l^2}{2m}\left(1\pm\sqrt{1-12m^2/l^2}\right)\;,
\end{equation}
where the asymptotic of the psedupotential $V_l(r)$ implies that for $12m^2/l^2<1$, the radial coordinate $r_+$ corresponds to the local minimum (the stable circular orbit) and $r_-$ corresponds to the local maximum of the pseudopotential (the unstable circular orbit). The remaining real solution exist for $12m^2/l^2=1$ where $r_+=r_-$, so in this case $r_\pm=6m$ is an inflection point. For $12m^2/l^2>1$, the local maximum of the pseudopotential does not exist, so all test particles moving towards the black hole are doomed to fall into it. Therefore, from now on we assume $l^2\geq 12m^2$.

For a given $E$ there exists a unique $l=l_{crit}$ such that the probe attains a minimum coordinate distance $r_{min}$ to a black hole without falling into it. When the probe is at the minimum distance, Eq.\ (\ref{maineq}) reduces to \mbox{$E=V_{l_{cirt}}[r_{min}=r_-(l_{crit})]$}. Using Eqs.\ (\ref{maineq},\ref{pot},\ref{Vp}) we have
\begin{equation}\label{V0}
2Er_{min}^3+2mr_{min}^2-l_{crit}^2r_{min}+2ml_{crit}^2=0\;.
\end{equation}
If the probe did not have any initial velocity at infinity, then $\gamma_\infty=e=1$ which implies $E=0$. In such a case, it is sufficient to subtract Eq.\ (\ref{Vp}) from Eq.\ (\ref{V0}) to find that $l_{crit}=r_{min}$. Substituting this result into Eq.\ (\ref{Vp}) or Eq.\ (\ref{V0}) shows that $r_{min}=4m$. The probe without propulsion which is non-relativistic at infinity can approach a black hole in the flyby at the coordinate distance not smaller than $r_{min}=4m$. In fact, we will show at the end of this subsection that $r_{min}$ might be approached only asymptotically with the probe endlessly spiraling towards an unstable circular orbit with $r=r_{min}$.

The case $E\neq 0$ needs separate attention. Multiplying Eq.\ (\ref{Vp}) by $2/3$ and subtracting Eq.\ (\ref{V0}) reduces the problem to the quadratic equation
\begin{equation}\label{lcrit}
l_{crit}^2=2\left(3E+\frac{2m}{r_{min}}\right)r_{min}^2\;.
\end{equation}
One may substitute $l_{crit}^2$ into Eq.\ (\ref{Vp}) which reduces to
\begin{equation}\label{eqrmin}
2Er_{min}^2+(1-6E)mr_{min}-4m^2=0\;.
\end{equation}
This equation has two real roots for
\begin{equation}
-\infty<E\leq-1/2\;\;\text{or}\;-1/18\leq E< 0\;\;\text{or}\;\ 0<E<+\infty\;.
\end{equation}
The condition $E\leq -1/2$ is inconsistent with $u^\alpha$ being a timelike vector, so we exclude it from our considerations. If $E=-1/18$, then the real root is degenerate. For $-1/18<E<0$ both roots are positive---they correspond to $r_-$ and $r_+$. For $E>0$, one of the roots is negative. We are interested in the positive root only, the one which corresponds to the local maximum of the pseudopotential. It is given for $-1/18\leq E<0$ or $0<E<+\infty$ by 
\begin{equation}\label{finsol}
r_{min}=\frac{m}{4E}\left(\sqrt{(1-6E)^2+32E}-(1-6E)\right)\;.
\end{equation}	
The critical value $E=-1/18$ corresponds to the innermost stable circular orbit $r_{min}=6m$ (see Fig.~\ref{fig0}).
The formula (\ref{finsol}) is valid also in the limit $E\rightarrow 0$ (the non-relativistic initial velocities of the probe) which gives $r_{min}\rightarrow 4m$ as expected. What is even more interesting, in the relativistic limit $E\rightarrow +\infty$ we obtain $r_{min}\rightarrow 3m$, thus we recover the size of the black hole shadow. In our setting, the condition $-1/18\leq E<0$ corresponds to the gravitationally bound probe which does not fall into the black hole. Our formulation of the problem excluded this case from direct considerations, but we point out that for such a probe there also exists a ``forbidden inner region'' with $r_{min}$ given in Eq.~(\ref{finsol}).

 The probe that is not gravitationally bound by a black hole may attain the minimum distance $r_{min}$ only asymptotically. In order to see that, consider the probe approaching $r_{min}$ from above with $l_{crit}$. We have
\begin{equation}
\frac{d\varphi}{dr}=\frac{d\varphi}{d\tau}\frac{d\tau}{dr}=\frac{u^\varphi}{u^r}=-\frac{l_{crit}}{r^2}\frac{1}{\sqrt{2[E-V_{l_{crit}}(r)]}}\;.
\end{equation}
Since $E=V_{l_{crit}}(r_{min})$, the derivative $d\varphi/dr$ blows up in the limit $r\rightarrow r_{min}$ and the probe rotate infinite number of times around the black hole. As pointed out by Hagihara \cite{hagihara}, it is a typical example of {\it cycles limites} which were introduced in the dynamical system theory by Poincaré. In order to avoid such a behavior, the gravitationaly unbound probe for a fixed energy per mass $e$ must have an angular momentum per mass at least slightly larger than the critical value $l_{crit}$. The image taken by the probe at an almost circular part of the orbit would be distorted as described in the article \cite{muller}.

To sum up, the closest possible non-destructive encounter of the propulsionless probe with a black hole should be planned by a distant observer as follows. The maximal value of the initial Lorentz factor $\gamma_\infty$ of the probe is set presumably by practical limitations. Its value defines the parameter $E=(\gamma_\infty^2-1)/2\geq 0$. For a given $E$, $r_{min}$ follows from Eq.~(\ref{finsol}) (a formula that is valid in the limits $E\rightarrow 0$, $E\rightarrow +\infty$).  Substituting $E$ and $r_{min}$ into Eq.~(\ref{lcrit}), we get $l_{crit}$ that defines a minimal angle at which the probe should be launched. The minimum coordinate distance $r_{min}$ corresponds to the unstable circular timelike geodesics and may be attained only asymptotically. The probe that flies away to infinity after a flyby must move with $l>l_{crit}$. 

\subsection{The shape of the orbits}\label{moredetails}

In Subsection \ref{sec:min}, we calculated the minimum distance to a black hole in a free flyby. The appropriate equations were derived within general relativity, but only simple algebraic manipulations of these equations were needed to obtain the main result. With a small amount of extra effort, one can find more details on the flyby. In this subsection, in order to keep the formulas as simple as possible, we consider a probe that is non-relativistic for distant observers ($\gamma_\infty=1$ which implies $e=1$ and $E=0$).

We describe the motion of our probe from the point of view of stationary observers in the Schwarzschild coordinates (\ref{metric}). The four-velocity $u_{O}^\alpha$ of stationary observers is given by 
\begin{equation}\label{uobs}
u_{O}^\alpha=\left(\frac{1}{\sqrt{1-2m/r}},0,0,0\right)\;,
\end{equation}
where the component $u^t_{O}$ follows from the normalization condition $u_{O}^\alpha {u_{O}}_\alpha=-1$. The Lorentz factor of the probe relative to these observers is given by
\begin{equation}\label{gammaobs}
\gamma(x^\mu)=-u^\alpha_{O}u_{\alpha}=\frac{e}{\sqrt{1-2m/r}}\;.
\end{equation}
Using the standard formula $\gamma=1/\sqrt{1-v^2}$, we find for $e=1$ the velocity of the probe relative to such observers $v=\sqrt{2m/r}$. A four-acceleration is defined as $a^\alpha=u^\beta\nabla_\beta u^\alpha$. The probe is moving on a geodesic, thus its four-acceleration vanishes. However, it is instructive to calculate the four-acceleration for stationary observers. Its single non-vanishing component is $a^r=\frac{m}{r^2}$. The magnitude 
\begin{equation}\label{acceleration}
a_O=|a|=\sqrt{a^\alpha a_\alpha}=\frac{1}{\sqrt{1-2m/r}}\frac{m}{r^2}
\end{equation}
gives the strength of the gravitational acceleration acting on stationary observers.

In general, geodesics in the Schwarzschild spacetime cannot be presented in an explicit form parametrized by $\tau$, $t$  or the affine parameter $\lambda$. For non-circular geodesics, it is possible to find explicit expressions in terms of a radial coordinate $r$. Such formulas, with a few exceptions, are complicated and make use of hypergeometric special functions or elliptic integrals \cite{chandra}. In special cases, like studies of a precession of a perihelion, one may assume that the test particle moves far away from the central body and some higher order relativistic terms may be neglected to obtain simple solutions. Since we would like to send a probe as close to a black hole as possible, one cannot use this kind of approximations along the full trajectory. Assuming $e=1$ simplifies formulas. Special functions are still needed in general, but not always as we show below. 

Let us consider the critical orbit $l=l_{crit}$ with $e=1$ (which implies $E=0$, $r_{min}=l_{crit}=4m$).
For the infalling probe $d\tau/dr=1/u^r=-[-2V_{l_{crit}}(r)]^{-1/2}$. Keeping in mind that
\begin{equation}
\begin{split}
\frac{dt}{dr}&=\frac{dt}{d\tau}\frac{d\tau}{dr}=\frac{1}{1-2m/r}\frac{1}{u^r}\;,\\
\frac{d\varphi}{dr}&=\frac{d\varphi}{d\tau}\frac{d\tau}{dr}=\frac{l}{r^2}\frac{1}{u^r}\;,
\end{split}
\end{equation}
and using knowledge how to calculate basic integrals we find the solution in the concise form
\begin{equation}
\begin{split}
\varphi(r)=&\sqrt{2}\ln\left(\frac{\sqrt{r}+2\sqrt{m}}{\sqrt{r}-2\sqrt{m}}\right)+\varphi_0\;,\\
\tau(r)=&-\frac{1}{3}\sqrt{\frac{2r}{m}}(r+12m)+4m\varphi(r)+\tau_0\;,\\
t(r)=&-\frac{1}{3}\sqrt{\frac{2r}{m}}(r+18m)+8m\varphi(r)+t_0\\
&-2m\ln\left(\frac{\sqrt{r}+\sqrt{2m}}{\sqrt{r}-\sqrt{2m}}\right)\;,
\end{split}
\end{equation}
where $\varphi_0$, $\tau_0$, $t_0$ are arbitrary constants. It follows from the form of the solution that the probe rotates an infinite number of times around a black hole as it approaches the minimum distance $r_{min}=4m$. In this sense, the probe is captured by a black hole. The minimally overcritical orbits $l=l_{crit}+\delta l$, where $\delta l\ll l_{crit}$, orbit a finite number of times. For these trajectories, at the point of the nearest approach $r=r_{min}+\delta r$, we have $u^r=0\Rightarrow V_{l_{crit}+\delta l}(r_{min}+\delta r)=0$. This leads to the quadratic equation that can be used to find relation between $\delta l$ and $\delta r$. We are interested only in the positive root. Substituting $r_{min}=4m$,
\begin{equation}
\delta r=\sqrt{\delta l}\left[\sqrt{\delta l}+\frac{4m+\delta l}{4m}\left(\sqrt{\delta l}+\sqrt{8m+\delta l}\right)\right]\approx \sqrt{8m\delta l}\;,
\end{equation}
which reveals that $\sqrt{\frac{\delta l}{4m}}/\frac{\delta r}{4m}\approx 1/\sqrt{2}$. One may expand $d\tau/dr$ in $\delta l$ and with a little effort find exact concise formulas for marginally overcritical orbits, but this expansion is valid only if $\sqrt{\delta l}/(r-4m)$ is small. Since near the black hole $r-4m$ approaches $\delta r$, the expression $\sqrt{\delta l}/(r-4m)$ converges to a positive non-zero constant $1/\sqrt{8m}$ and such an approximation cannot be used in general near the black hole. Of course, for a fixed $\delta l$ one may expand $d\tau/dr$ in $r$ about $r_{min}+\delta r$ and find exact solutions near the circular orbit $r_{min}=4m$. However, for the purposes of this article, it is more convenient to determine marginally overcritical orbits numerically.

\begin{figure}[t!]
\centering
\includegraphics[scale=0.5]{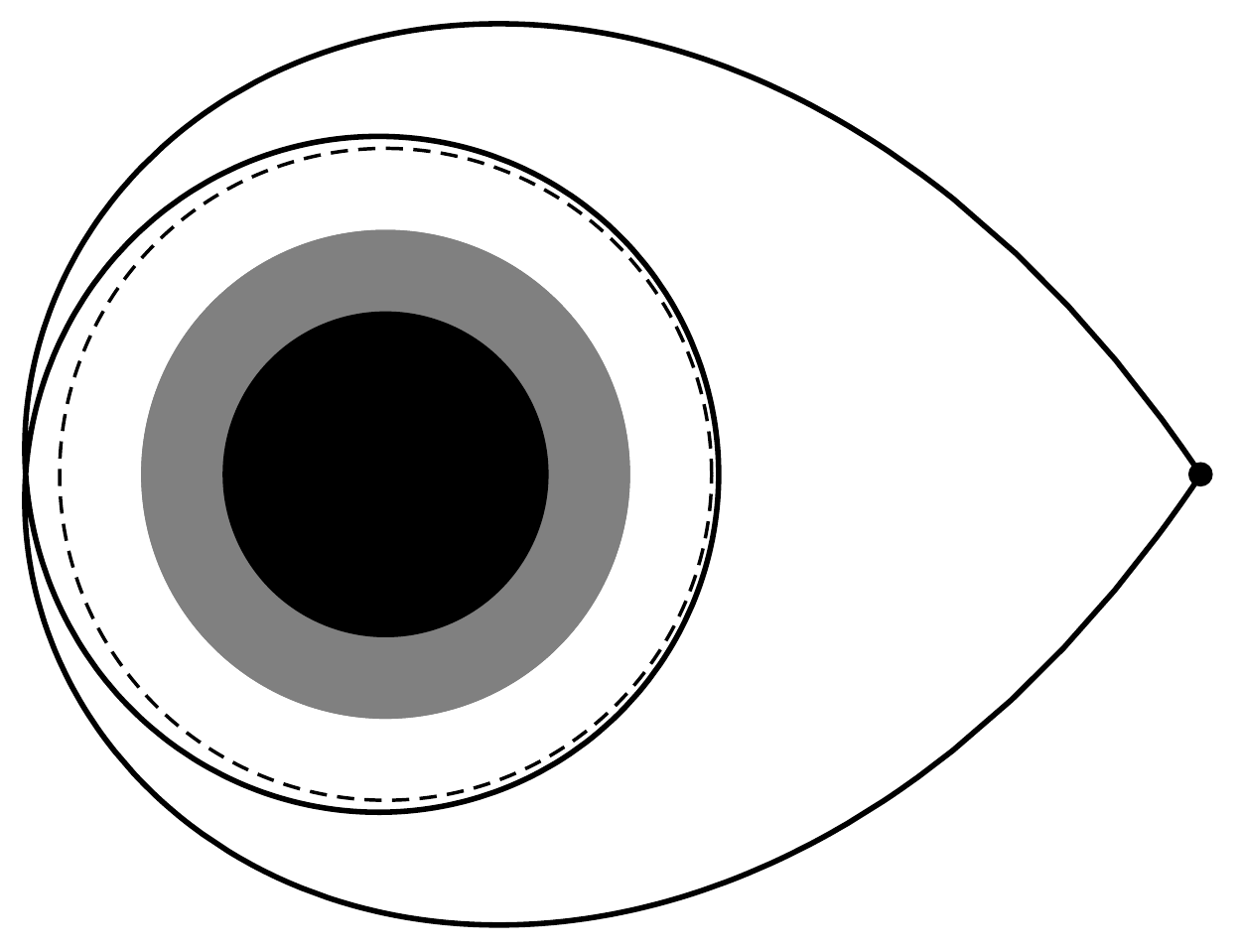}
\caption{The ``eye-shaped'' marginally overcritical timelike geodesic. The black hole corresponds to a black disk with a radius $r=2m$. The larger gray disk has a radius $r=3m$ which corresponds to the photon's unstable circular orbit. The orbit $r=4m$ is indicated as a dashed circle. For clarity of presentation, the probe starts at $r=10m$. Its initial and final position is marked with a dot.}
\label{fig1}
\end{figure}

One may think of many interesting types of marginally overcritial geodesics that can be useful in black hole exploration. We focus on the ``eye-shaped'' trajectory. Far away from the black hole a stationary observer sends a probe that collects data and returns to the starting position. The example of such trajectory is presented in Fig.~\ref{fig1}.

The set of equations that is solved numerically has the form
\begin{equation}
\begin{split}
\frac{dt}{d\tau}=&\frac{1}{1-\frac{2m}{r}}\;,\\
\frac{d^2r}{d\tau^2}=&-\frac{d}{dr}\left[V_{4m+\delta l}(r)\right]\;,\\
\frac{d\varphi}{d\tau}=&\frac{4m+\delta l}{r^2}\;,
\end{split}
\end{equation}
where $dr/d\tau$ was differentiated to avoid numerical problems with the square root. The initial conditions are $t(0)=0$, $r(0)=r_O=50m$, $\varphi(0)=0$, and $\frac{dr}{d\tau}(0)$ is calculated from the normalization condition $u_\alpha u^\alpha=-1$. We assume that the probe returns to its initial position after going once around the black hole. This condition sets $\delta l$ to some small value $\delta l \ll 4m$ which may be determined numerically by the bisection method applied to initial conditions. The problem formulated in this way is scale-invariant, so without loss of generality one may substitute $m=1$ and reintroduce the physical units after integration (for a given $m$). We determined $\delta l\approx 0.005556m$. For the initial conditions chosen, we have found the duration of the maneuver, as measured by the clocks moving with the probe, $\Delta\tau\approx 433.8m$, and as measured by an observer at infinity, $\Delta t\approx 507.3m$. (We calculate these intervals for three astrophysical black holes in Table~\ref{tab1}.) From Eq.~(\ref{uobs}), we know that for the stationary observer who sends the probe at $r_O=50m$ the time passes at a different rate than for observers at infinity, namely, $dt/d\tau_O=u^t_{O}=\frac{5}{2\sqrt{6}}$ which gives $\Delta \tau_O\approx 497m$. To maintain a stationary position relative to a black hole, such an observer must overcome its acceleration which is given by Eq.~(\ref{acceleration}) and at $r_O=50m$ equals $a_0=|a|=1/\sqrt{6}\times 10^{-3}m^{-1}$. In our setting, the probe is not falling from infinity, but it is released at $r_O=50m$. A non-zero initial velocity of our probe equals to the velocity it would gain in a free fall from infinity (with a zero initial velocity at infinity). Using the formula $v=\sqrt{2m/r}$ which follows from Eq.~(\ref{gammaobs}) we find $v_{O}=0.2c$, where $c$ denotes the speed of light. The maximum velocity relative to the local stationary observers is obtained at the closest approach to a black hole at $r=4m+\delta r$ where $\delta r$ may be calculated in terms of $\delta l$ as in Sec.~\ref{moredetails}. We find $v_{max}\approx 0.69c$.


\subsection{M87*, Sagittarius A*, Cygnus X-1}\label{examples}

In order to gain intuition into a black hole flyby, we evaluate characteristic parameters of the orbit for masses of three astrophysical black holes. We consider two supermassive black holes, M87* in the center of the galaxy M87, Sagittarius A* in our Milky Way, and the stellar-size black hole Cygnus X-1 \cite{footnote2}. For obvious reasons, we ignore the question of how to get spacecraft in close proximity ($r_O=50m$) to these black holes. (What is close proximity in terms of cosmic distances is considered to be ``far away'' in terms of spacetime curvature deviation from flatness.)

We apply dimensional analysis to change from/to geometric units $G=c=1$. A conversion factor is given by an appropriate combination of 
$G$ and $c$ in standard SI units.
As an example consider two quantities: a time interval and an acceleration. In the geometric units these quantities are given in terms of a black hole mass and an inverse of the mass, respectively. The mass is measured in meters, but we usually know it in kilograms. Thus, the mass in kilograms has to be multiplied by $G/c^2$ to obtain the mass in meters which is required by our formulas. The results ($\sim m$ and $\sim 1/m$) have to be multiplied by the conversion factors $1/c$ and $c^2$ for a time interval and an acceleration, respectively. The intermediate step (kilograms to meters conversion) may be skipped and then the appropriate conversion factors are $G/c^3$, $c^4/G$.

We have summarized the parameters of orbits in Table~\ref{tab1}. The table reveals technological challenges facing this hypothetical mission and shows that black holes are, indeed, the extreme objects. The gravitational attraction at $r_O=50m$ for the Sgr A* and Cyg X-1 black holes is definitely too large for any spacecraft to maintain stationary position there with present and imaginable future propulsion technology. For the M87* black hole $a_O=0.39\;g$. This does not look like a lot. However, if we assume that such an acceleration must be kept for $184$ days (the observer remains stationary for the duration of the mission), then it might be highly problematic. The supermassive black holes M87*, Sgr A* are more ``astronaut friendly,'' but in case of the stellar black hole Cyg X-1, the tidal forces are deadly for humans already at $r_O=50m$. In our calculations, we assumed that the probe is released at $r_O=50m$ with initial velocity $v=0.2c$. Such relativistic initial speeds for gram-scale robotic spacecrafts are imaginable in the foreseeable future as the project {\it Breakthrough Starshot} shows \cite{starshot,marx}. The ``eye-shaped'' trajectories desynchronize the proper times of distant observers and the probe. The time dilation is noticeable, but we point out that this kind of effects are usually much more severe for rotating black holes (e.g.,\ trajectories approaching the innermost stable circular orbits for a near extreme Kerr black hole which were studied in the article \cite{rutkowski}). We note that for the orbit studied, the time duration of the travel around the Cyg X-1 black hole is too short to be perceived visually by a human being. Finally, we mention that for the Sgr A* black hole the radius $r_O=50m$ corresponds to around two astronomical units, so the considered trajectory is easy to imagine in the Solar System scale. Similarly, the appropriate scale for the Cyg X-1 black hole is around $1107\;km$, which corresponds to the size of a typical European country. This scale with a characteristic time interval measured in milliseconds sets an impressive vision of a macroscopic dynamical system near a stellar-size black hole where all interesting things happen in a blink of an eye.

\section{Summary}

The structure of timelike geodesics near the Schwarzschild black hole is well-known. In this article, we brought to light, in a way that is useful for teaching elementary general relativity, a particular aspect of this structure, namely, the minimum distance at which a black hole can be approached by a test massive particle that is freely falling from infinity. We showed that this problem is analogous to the problem of null geodesics and the black hole shadow. Photons generated outside of the sphere $r=3m$ and received by a distant observer did not cross the surface $r=3m$. (Of course, photons emitted above the event horizon $r=2m$ and inside/at the sphere $r=3m$ still can approach a distant observer.) A similar impenetrable boundary exists for free, massive, gravitationally unbound particles approaching a black hole from a distance. The size of this boundary depends on the energy per mass of the particle as measured by an observer at infinity and lies in the interval $3m<r\leq 4m$ where $3m$ corresponds to the ultra-relativistic particles and $4m$ corresponds to non-relativistic particles. In other words, the analog of the black hole shadow for free, massive, unbounded gravitationally particles that are non-relativistic far away from a black hole corresponds to $4m$. It shrinks smoothly to the photon limit $3m$ as the kinetic energy of test massive particles increases. For gravitationally bounded particles that do not fall into the black hole a forbidden region also exists. The minimum distance at which they can approach a black hole depends on their energy per mass relative to some local observers and lies in the interval $[4m,6m]$, where $6m$ corresponds to the innermost stable circular orbit. Thus, for gravitationally bound and unbound free particles of given energy per mass that originated away from a black hole, a boundary of the forbidden region (bracketed by $r=3m$ and $r=6m$) cannot be smaller than the radius of the innermost admissible circular orbit. We found the concise analytic formula for the critical geodesic that approaches $r=4m$ asymptotically. 

The problem studied in this article is formulated as a flyby of a probe near a Schwarzschild black hole. It can be explored in many ways, an example of which is provided in our work. In addition to the main results described above, we investigated in detail a particular orbit. Its physical properties have been found for three astrophysical black holes: M87*, Sagittarius A*, and Cygnus X-1.

\begin{table*}[t] 
\centering 
\caption{The parameters of the ``eye-shaped'' flyby [as in Fig.~(\ref{fig1})] which starts at $r_O=50m$. The $r=2m$ corresponds to the size of the event horizon. The parameter $a_O$ is an acceleration of a stationary observer at $r_O=50m$ needed to overcome gravity of the black hole. The time intervals $\Delta\tau$, $\Delta\tau_O$, $\Delta t$ are measured by clocks moving with a probe, at $r_O=50m$, at infinity, respectively. $M_\odot=1.988435\times 10^{30}\;kg$ denotes the mass of the Sun, $g=9.80665\;m/s^2$ is the standard acceleration due to the Earth's gravity, $au$ are astronomical units.} 
\begin{ruledtabular} 
\begin{tabular}{l c c c c c c c} 
Black hole & m & $r=2m$ & $r_O=50m$ & $a_O$ & $\Delta\tau$ & $\Delta\tau_O$ & $\Delta t$ \\ 
\hline 
M87* & $6.5\times 10^9 M_\odot$ & $128\; au$ & $3208\; au$ &$0.39\;g$ & $161\; days$ & $184\;days$ & $188\; days$ \\ 
Sgr A* & $4.15\times 10^6 M_\odot$ & $0.08\; au$ &$2 \; au$ & $610\; g$  & $2\;h\;28\;min$ & $2\;h\;49\;min$ & $2\;h\;52\;min$ \\ 
Cyg X-1 & $15 M_\odot$ & $44.3\; km$ & $1107 \; km$& $1.7\times 10^8\;g$ & $3.2\;ms$ & $3.67\;ms$ & $3.74\;ms$ \\ 
\end{tabular} 
\end{ruledtabular} 
\label{tab1} 
\end{table*}


\begin{thebibliography}{99}
\bibitem{footnote1} Those readers who are familiar with the concept of the Killing vectors may apply here more general reasoning. The scalar products of the Killing vectors $\partial_t$, $\partial_\varphi$ and the probe four-velocity is conserved along the geodesic.
\end{thebibliography}
\end{document}